\documentclass[]{interact}

\usepackage{epstopdf}
\usepackage[caption=false]{subfig}

\usepackage[numbers,sort&compress]{natbib}
\bibpunct[, ]{[}{]}{,}{n}{,}{,}
\renewcommand\bibfont{\fontsize{10}{12}\selectfont}

\theoremstyle{plain}

\theoremstyle{definition}

\theoremstyle{remark}

\begin{document}

\articletype{ARTICLE - arXiv.org}

\title{How much does it cost to complete a sticker collection? A simulation study using R}


\author{
	\name{Edson Zangiacomi Martinez\textsuperscript{a}\footnote{Corresponding author: Ribeir\~{a}o Preto Medical School, Av. Bandeirantes 3900, University of S\~{a}o Paulo (USP), Ribeir\~{a}o Preto, 14049-900, Brazil. E-mail: edson@fmrp.usp.br}, Jorge Alberto Achcar\textsuperscript{a}, Marcos Vinicius de Oliveira Peres\textsuperscript{b}, Ricardo Puziol de Oliveira\textsuperscript{c}, Regina Albanese Pose\textsuperscript{d}}
	\affil{\textsuperscript{a}Faculdade de Medicina de Ribeir\~{a}o Preto, Universidade de S\~{a}o Paulo (USP), Ribeir\~{a}o Preto, Brazil.\\ 
		   \textsuperscript{b}Universidade Estadual do Paraná, Campus de Paranava\'{\i}, Brazil.\\ 
		   \textsuperscript{c}Universidade Estadual de Maring\'{a}, Maring\'{a}, Brazil.\\
		   \textsuperscript{d}Universidade Municipal de S\~{a}o Caetano do Sul (USCS), S\~{a}o Caetano do Sul (SP), Brazil.}}

\maketitle

\begin{abstract}
This paper considers some computational issues related to the problem of the coupon collector behavior. In addition, we propose a simulation algorithm to find the solution to the problem, where we present as a numerical illustration an album of stickers related to the Qatar 2022 Official Football World Cup, which is being sold in Brazilian newsstands. In this case, the complete album has $N = 670$ stickers, which are sold in packages containing $n = 5$ stickers each. The simulation was based on a computational code written in R. For this application we obtained as a solution, an average number of $950$ packages needed to complete the album, at an average cost of $3,800$ reals (Brazilian currency), assuming that the collector does not exchange any of its repetitions with nobody. All computer code used in this article is provided on GitHub. The assumed simulation study could be useful for statistics teachers to motivate their students for lectures in computational statistics and fundamentals of probability.
\end{abstract}


\section{Introduction}

The coupon-collector's problem is a classical probability problem. It considers one person that collects coupons and assumes that there is a finite number of different coupons, labeled by the numbers $1, 2, \dots ,N$. In this case, we may be interested to find the probability of completing the collection after acquiring exactly $n$ ($n > N$) coupons or in the estimation of the mean number of coupons needed to complete the collection. It is believed that this problem appeared earlier in the work \textit{De Mensura Sortis} (On the Measurement of Chance), written by Abraham de Moivre in 1711, and further results were obtained by Laplace and Euler (\citeauthor{Ferrante2014}, \citeyear{Ferrante2014}). It has many applications in different areas of science such as ecology and evolution (\citeauthor{Zoroa2014}, \citeyear{Zoroa2014}), computer science (\citeauthor{Nakata2006}, \citeyear{Nakata2006}), engineering (\citeauthor{Boneh1997}, \citeyear{Boneh1997}) and quality engineering (\citeauthor{Luko2009}, \citeyear{Luko2009}).

This article revisits the coupon-collector's problem and introduces a simulation program developed in the R language. For this purpose, we consider as a special illustrative example, the official 2022 Qatar soccer World Cup sticker album, released in Brazil in August 2022, which is being sold in Brazilian news-stands. The results obtained in this study, especially the results related to the simulation study, could be very useful for statistics teachers to motivate students in teaching fundamentals of probability and statistics in basic and intermediate courses.

\section{Methods}

\subsection{The sticker-collector's problem}

Let us consider a population whose members are of $N$ different types (in our case, all stickers from a sticker book). For $1\leq j\leq N$, let us denote by $p_{j}$ the probability that a sticker is of type $j$, where $p_{j}>0$ and $\sum \nolimits_{j=1}^{N}p_{j}=1$. The simplest case occurs when one takes%
\[
p_{1}=p_{2}=...=p_{N}=\frac{1}{N},
\]%
or say, each sticker is produced in equal numbers and so has an equal chance
of being found in the next packet that the collector open. In this case, $%
R_{i}$ denotes the additional number of stickers that the collector needs to
purchase to pass from $i-1$ to $i$ different types of stickers in her/his
collection. Note that $R_{i}$ has a geometric law with parameter%
\[
q_{i}=\frac{N-(i-1)}{N}.
\]

Therefore, $q_{1}=1$, $q_{2}=\frac{N-1}{N}$, $q_{3}=\frac{N-2}{N}$, and so
on. If $R$ is the number of stickers that the collector has to buy in order
to find all $N$ existing different stickers, we have that%
\[
R=R_{1}+R_{2}+...R_{N},
\]%
and%
\begin{eqnarray}
	E(R) &=&\sum\limits_{i=1}^{N}E(R_{i})=\sum\limits_{i=1}^{N}\frac{1}{q_{i}}%
	=\sum\limits_{i=1}^{N}\frac{N}{N-(i-1)} \nonumber \\
	&=&N\sum\limits_{i=1}^{N}\frac{1}{N-(i-1)}=N\sum\limits_{i=1}^{N}\frac{1}{i}. \label{ET1}
\end{eqnarray}%
\bigskip 

An approximation of (\ref{ET1}) is given by%
\begin{equation}
E(R)\approx N(\ln N+\gamma )+0.5, \label{apx}
\end{equation}%
where $\gamma \approx 0.5772156649$ is the Euler-Mascheroni constant
(\citeauthor{Doumas2015}, \citeyear{Doumas2015}), defined as the limit of the sequence%
\[
\gamma =\lim_{N\rightarrow \infty }\left( \sum\limits_{i=1}^{N}\frac{1}{i}-\ln N\right).
\] 

\subsection{The collector's problem in the case of sampling in groups of constant size}

We now formulate the collector's problem as follows: given a collection with $N$ different stickers, how often must we draw subsets of $1\leq n\leq N$ distinct stickers in order to see each stickers at least once? This subsection is based on \citeauthor{Stadje1990} (\citeyear{Stadje1990}), \citeauthor{Leite1992} (\citeyear{Leite1992}), \citeauthor{Adler2001} (\citeyear{Adler2001}), and \citeauthor{Diniz2016} (\citeyear{Diniz2016}), and considers that the stickers are sold in packets of $n$ stickers each. Let $N$ be the total number of stickers in a collection and $T$ denotes the number of packets necessary to complete the collection. The probability that more than $k$ packets are needed to complete the collection is given by $P(T>k)$. Let us consider the following event: 
\[
A_{i,k}:\text{ sticker }i\text{ is not selected in the first }k\text{ packets.}
\]

The probability $P(T>k)$ is thus given by%
\begin{equation}
	P(T>k)=P\left( \bigcup\limits_{i=1}^{N}A_{i,k}\right).  \label{P(T>k)}
\end{equation}

Assuming independence between packets, we have $P\left( A_{i,k}\right)
=p_{1}^{k}$ where%
\[
p_{r}=\frac{\binom{N-r}{n}}{\binom{N}{n}}
\]%
is the probability that no sticker from a subset of $r$ stickers is found in
the $j$-th packet with $n$ stickers each ($k\geq 1$, $n<N$, $N-\min
(kn,N)\leq r\leq N-n$). Diniz et al. (2016) observed that $p_{r}$ neither
depends on the particular subset nor the particular packet $j$ and, for
distinct elements $i_{1}$, $i_{2}$, ..., $i_{s}$ of the set $\{1,...,N\}$,
the authors used the notation%
\begin{eqnarray}
	p_{1}^{k} &=&P(A_{i_{1},k})  \nonumber \\
	p_{2}^{k} &=&P(A_{i_{1},k}\cap A_{i_{2},k})  \nonumber \\
	&&\vdots   \nonumber \\
	p_{s}^{k} &=&P(A_{i_{1},k}\cap A_{i_{2},k}\cap ...\cap A_{i_{s},k}).
	\label{pk}
\end{eqnarray}

From (\ref{P(T>k)}) and (\ref{pk}), we have%
\begin{eqnarray}
	P\left( T>k\right) 
	&=&\sum\limits_{i_{1}=1}^{N}P(A_{i_{1},k})-\sum\limits_{i_{1}\neq
		i_{2}}P(A_{i_{1},k}\cap A_{i_{2},k})+... \nonumber \\
	&&+(-1)^{N-1}P(A_{1,k}\cap A_{2,k}\cap ...\cap A_{N,k}) \nonumber \\
	&=&\binom{N}{1}p_{1}^{k}-\binom{N}{2}p_{2}^{k}+...+(-1)^{N-1}\binom{N}{N}%
	p_{N}^{k} \nonumber \\
	&=&\sum\limits_{i=1}^{N}(-1)^{i-1}\binom{N}{i}p_{i}^{k},\text{ \ } \label{PTmK}
\end{eqnarray}%
and%
\begin{eqnarray*}
	P\left( T=k\right)  &=&P(T>k-1)-P(T>k) \\
	&=&\sum\limits_{i=1}^{N}(-1)^{i-1}\binom{N}{i}p_{i}^{k-1}-\sum%
	\limits_{i=1}^{N}(-1)^{i-1}\binom{N}{i}p_{i}^{k} \\
	&=&\sum\limits_{i=1}^{N}(-1)^{i-1}\binom{N}{i}\left(
	p_{i}^{k-1}-p_{i}^{k}\right)  \\
	&=&\sum\limits_{i=1}^{N}(-1)^{i-1}\binom{N}{i}p_{i}^{k-1}\left(
	1-p_{i}\right) .
\end{eqnarray*}

The expected number of packets necessary to complete the collection is given by%
\begin{equation}
	E(T)=\sum\limits_{i=1}^{N}(-1)^{i-1}\binom{N}{i}\frac{1}{1-p_{i}}.
	\label{ET}
\end{equation}

\subsection{A computational simulation}

We use the R software to simulate the number of sticker packets that are required to complete one collection sequentially. If we run this simulation a relatively large number of times, i.e., assuming many collectors are buying stickers simultaneously, we can estimate the average number of packets necessary to fill the albums. Considering the official 2022 Qatar soccer World Cup sticker album, we have $N=670$ and $n=5$. We simulate each purchased packet of stickers as a uniform sample of size $n$ drawn from a population of size $N$ with replacement. Given $N$ different stickers, the probability of finding $n$ different stickers in a packet of $n$ stickers is
\[
\frac{N(N-1)\cdots (N-n+1)}{N^{n}}=\prod\limits_{j=1}^{n-1}\left( 1-\frac{j}{N}\right).
\]
For $N=670$ and $n=5$ we have
\[
\prod\limits_{j=1}^{4}\left( 1-\frac{j}{670}\right) \simeq 0.985.
\]%
The steps of the algorithm are as follows.
\begin{enumerate}
	\item Assume that $N=670$ is the total in the album, $k = 5$ is the number of stickers in a packet, $H = 0$ is the number of stickers that the collector have, and $D = 0$ is the number of duplicate stickers.
    \item Let \texttt{is.missing} be a vector with the labels of the missing stickers. Initially \texttt{is.missing} is given by $\left\lbrace 1,2,3,...,670 \right\rbrace $.
    \item Let \texttt{stickers.b} $= 0$ be the number of packets that are required to complete the collection.
    \item Simulate one packet by taking a random sample of size $k = 5$ from the elements of $\left\lbrace 1,2,3,...,670 \right\rbrace $.
    \item Verify that each of the simulated stickers belongs to the vector \texttt{is.missing}. If a sticker does not belong to this vector, it is a duplicate sticker. In this case we take $D = D + 1$. Otherwise we exclude the sticker from the vector \texttt{is.missing} and take $H$ equal to $N$ minus the number of elements of \texttt{is.missing}.
    \item Take \texttt{stickers.b} $=$ \texttt{stickers.b} $+1$.
    \item Repeat the items (4) to (6) while $H<N$.
\end{enumerate}

In order to provide a didactic approach to this simulation, a code in R was developed to help visualize in an animation the steps of the algorithm until an album is filled.

We can use this algorithm to simulate a large $B$ number of collections, and from there we can estimate the average number of packets necessary to fill the albums without using the equations (\ref{ET1}), (\ref{apx}), or (\ref{ET}).

\subsection{Computational details}

All codes and data underlying this work are available in GitHub at https://github.com/edsonzmartinez/stickers.

\section{Results}

According to the expression (\ref{ET1}), an average of $E(R) = 4,748$ stickers are required to complete a collection of $N=670$ stickers. We get the same result for $E(R)$ if we use the approximation given by expression (\ref{apx}). Alternatively, considering $N=670$ and that the stickers come in packets of $n=5$, we approximated the expression (\ref{ET}) to be 946.98 using the WolframAlpha calculator (\citeauthor{Dimiceli2010}, \citeyear{Dimiceli2010}). Rounding up, we get 947 packets and $947 \times 5 = 4,735$ stickers. At a cost of $4$ Brazilian reals a packet, this implies a total average cost of $947 \times 4 = 3,788$ Brazilian reals (BRL) per album (as of August 2022, 1 BRL was equivalent to 0.19 US dollars).

\begin{figure}[h!]
	\centering
	\includegraphics[width=1\linewidth]{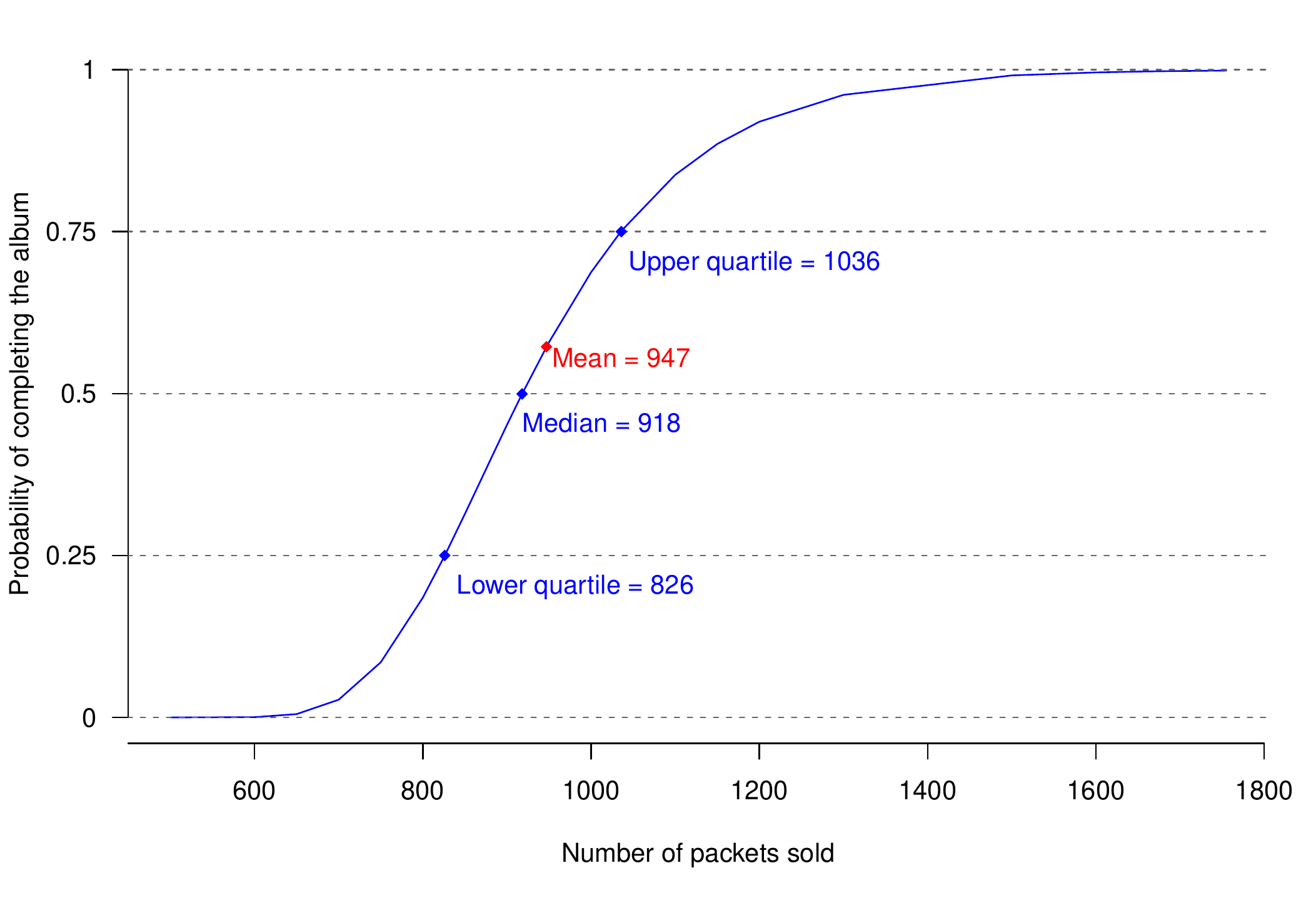}
	\caption[figDES]{Probability of completing the album according to the number of packets sold.\bigskip}
	\label{fig:Fig01}
 \end{figure}

Table \ref{tab01} shows the probability $1 - P(T>k)$ of completing the album given the number $k$ of purchased packets and the corresponding cost in BRL. Values of $P(T>k)$ were obtained using the expression (\ref{PTmK}). In addition, Figure \ref{fig:Fig01} describes the  relationship between $1 - P(T>k)$ and $k$. Table \ref{tab01} and Figure \ref{fig:Fig01} show that the probabilities of completing the album by purchasing 826, 918 and 1036 packets are, respectively, near to 0.25, 0.50, 0.75 (the lower quartile, median, and upper quartile). If a collector purchases 1500 packets, he/she will have a chance of about 0.99 to complete the album at a cost of 6,000 BRL.

\begin{table}[h!]
	\renewcommand{\arraystretch}{0.85}
	\caption{Probability of completing the album given the number $k$ of purchased packets.}
	\label{tab01}
	\begin{center}
		\tabcolsep=0.135cm
		\begin{tabular}{ccc}
			\hline \hline \noalign{\smallskip}
			$k$ & $1 - P(T>k)$ & Cost (BRL) \\
			\noalign{\smallskip}\hline\noalign{\smallskip}
500  & 0.00001 & 2000.00 \\
600  & 0.00044 & 2400.00 \\
650  & 0.00516 & 2600.00 \\
700  & 0.02735 & 2800.00 \\
750  & 0.08516 & 3000.00 \\
800  & 0.18488 & 3200.00 \\
826  & 0.24972 & 3304.00 \\
850  & 0.31416 & 3400.00 \\
900  & 0.45170 & 3600.00 \\
918  & 0.49950 & 3672.00 \\
950  & 0.57939 & 3800.00 \\
1000 & 0.68734 & 4000.00 \\
1036 & 0.75114 & 4144.00 \\
1100 & 0.83773 & 4400.00 \\
1150 & 0.88541 & 4600.00 \\
1200 & 0.91974 & 4800.00 \\
1300 & 0.96123 & 5200.00 \\
1500 & 0.99120 & 6000.00 \\ 
1600 & 0.99583 & 6400.00 \\
1650 & 0.99713 & 6600.00 \\
1700 & 0.99803 & 6800.00 \\
1750 & 0.99864 & 7000.00 \\
1755 & 0.99869 & 7020.00 \\  \noalign{\smallskip} \hline \hline	
		\end{tabular}
\end{center}
\end{table} 

\subsection{Simulation results}

\begin{figure}[h!]
	\centering
	\includegraphics[width=1\linewidth]{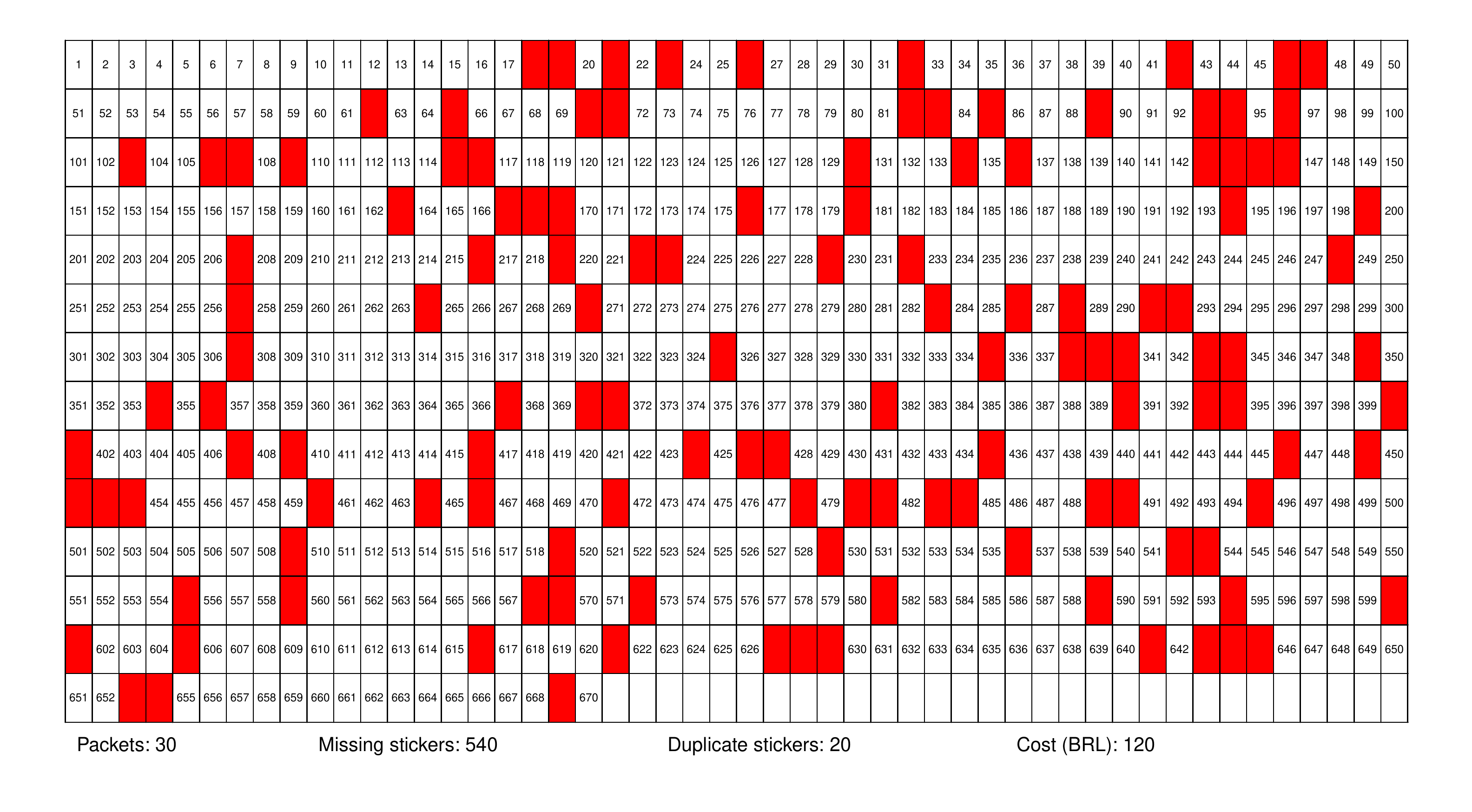}
	\caption[figDES]{Results of a simulation after the purchase of 30 sticker packets.\bigskip}
	\label{fig:Fig02}
\end{figure}

\begin{figure}[h!]
	\centering
	\includegraphics[width=1\linewidth]{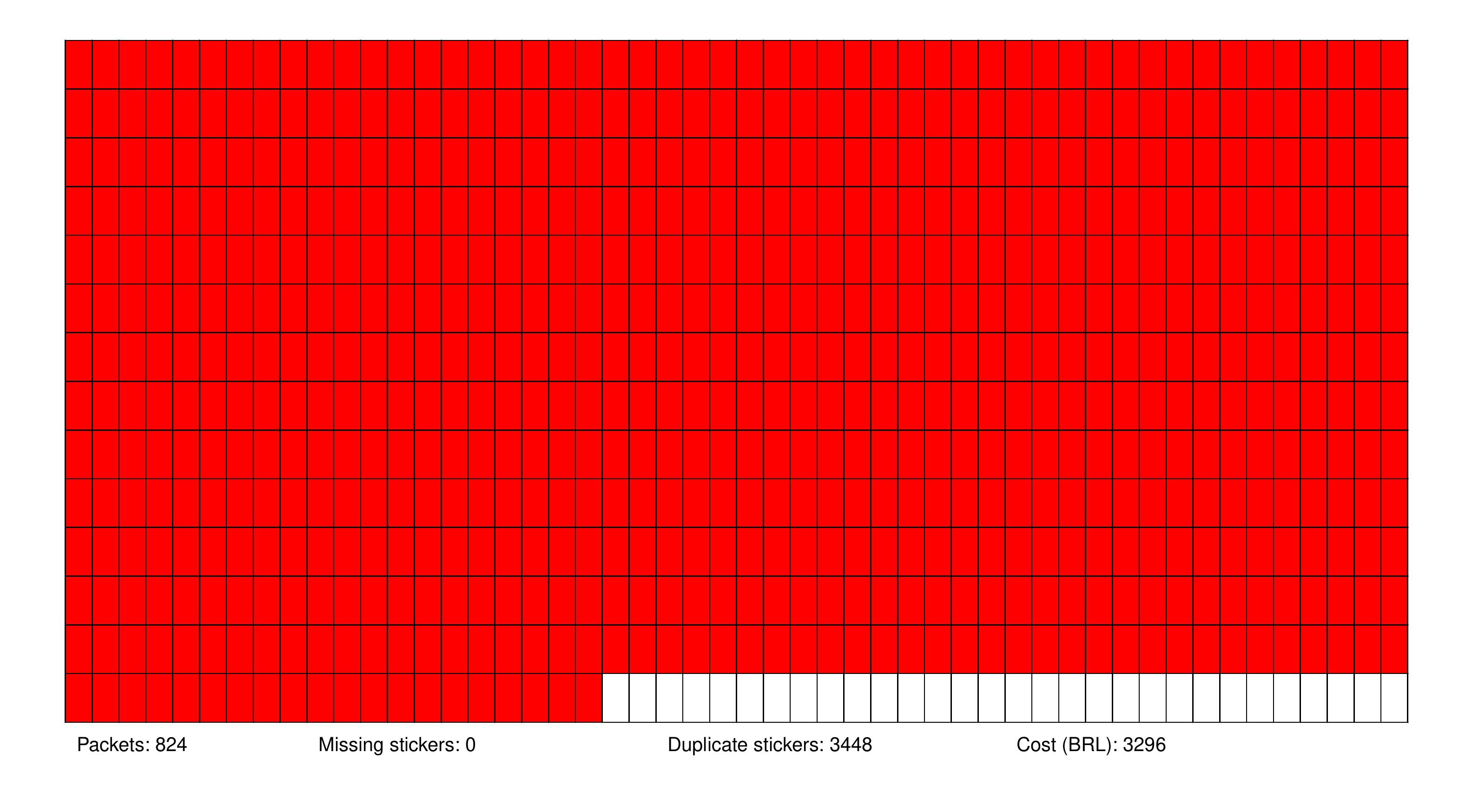}
	\caption[figDES]{Final results of the simulation, where it was required 824 packets to complete the album.\bigskip}
	\label{fig:Fig03}
\end{figure}

\begin{figure}[h!]
	\centering
	\includegraphics[width=1\linewidth]{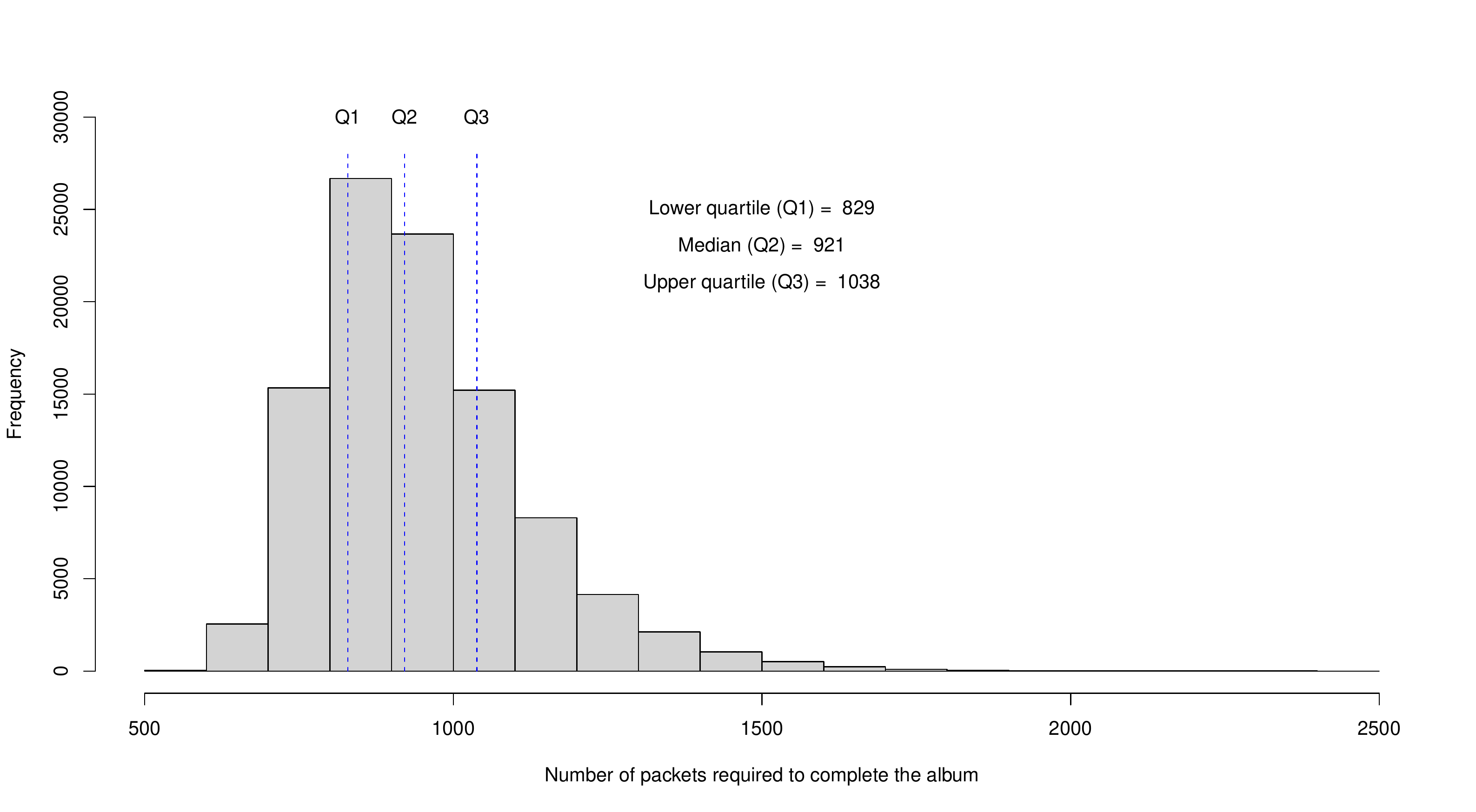}
	\caption[figDES]{Distribution of the number of packets required to complete $B = 100.000$ independent collections.\bigskip}
	\label{fig:Fig04}
\end{figure}

Figure \ref{fig:Fig02} shows a grid representing the results of a simulation after the purchase of 30 sticker packets. Continuing this simulation, Figure \ref{fig:Fig03} shows that it was required 824 packets to complete the album, at a cost of 3,296 BRL. Based on this algorithm, we simulated $B = 100.000$ independent collections. The number of required packets ranged from 547 to 2,455, and the corresponding cost ranged from 2,188 to 9,820 BRL. Figure \ref{fig:Fig04} shows a histogram of the distribution of the number of packets required to complete these 100.000 collections. The average number of required packets was 950 at a cost of 3,800 BRL. The lower quartile, median, and upper quartile were 829, 921, and 1,038, respectively, at corresponding costs of 3,316, 3,684, and 4,152 BRL. Therefore, the results of our simulation are close to that showed in Table \ref{tab01} and Figure \ref{fig:Fig01}, based on the expression (\ref{PTmK}).

\section{Discussion}

Using a simulation study, we show that it is necessary to purchase on average $950$ packets of $n = 5$ stickers, at a cost of 3,800 BRL, to complete a collection of $N = 670$ stickers. Similar results were obtained using expression (\ref{ET}), which does not require the use of computational simulation. However, these results ignore the facility for buying missing stickers and that collectors can swap stickers with other collectors. Intuitively, collectors know that filling a collection is easier when they exchange duplicated stickers with other people. According to \citeauthor{Hayes2006} (\citeyear{Hayes2006}), although statisticians frequently use the theory of the coupon collector's problem to respond to inquiries from newspapers about the cost of filling sticker albums, it is possible that the only way to find the true answer is by simulating the trading environment of the school playground (or other social environments). In this way, authors such as \citeauthor{Braband2015} (\citeyear{Braband2015}), \citeauthor{Morgado2019} (\citeyear{Morgado2019}), and  \citeauthor{Rosa2021} (\citeyear{Rosa2021}) proposed different extensions of the coupon-collector's problem considering that collectors trade duplicate stickers with other people. These articles show how trading repeated stickers with other collectors reduces substantially the amount of stickers that the collector needs to complete the album.

Although the coupon-collector's problem contextualized in collections soccer championships' stickers has already been extensively explored in the literature, the present article can contribute to educating students about some fundamentals of probability and computational statistics by providing detailed R codes for implementing the problem. We believe that the simulation study presented in this article can benefit high school and college students. For example, by showing that the number of sticker packets needed to complete the collection is not a fixed number but an amount that can vary from one collector to another, we exemplify the concept of a random phenomenon. It is possible that one collector will complete the collection earlier than another, but this is determined by chance.

\end{document}